# Inter-basin fluctuations in glass-forming liquids: short time and long time thermodynamic susceptibilities

Alexander Z. Patashinski, and Mark A. Ratner

Department of Chemistry and Materials Research Center, Northwestern University

Evanston, IL 60208

## Abstract

The thermodynamic effects of local structure fluctuations in glassformers are analyzed in terms of energy basins and inter-basin hopping. Depending on the time-scale of measurement, one observes short-time thermodynamic properties related to a narrow set of basins, or equilibrium properties that include structural relaxation. The inter-basin hopping is manifested by the fluctuations of the pressure, internal energy, and other thermodynamic short time characteristics. Formulas relating the inter-basin fluctuations of pressure and internal energy to the differences between the long- and short-time susceptibilities are found. Based on obtained relations, we discuss the relative sensitivity of the structure to temperature and pressure.

## 1. The physical picture of a supercooled liquid

Extensive study[1,2] has revealed common qualitative features in the cooling and supercooling behavior of glassformers but also substantial differences in the details of this behavior (see [3,4,5,6,7,8,9,10,11] and references therein). Upon cooling and supercooling, both the viscosity η and structural relaxation time θ increase exponentially (strong glassformers) or faster (fragile glassformers). In the current literature, this behavior is explained, following the ideas of Goldstein[12] and Stillinger (see [13] and references therein), by a special topography of the potential energy landscape in the part of the 3N-dimensional configuration space of a glassformer occupied by the amorphous state. Here, N is the macroscopically large number of particles in the system. The energy landscape picture is widely used in the literature[14,15,16,17,18,19], to explain many static and dynamic properties of supercooled liquids. It is assumed that the energy landscape of a liquid may be described in terms of energy basins separated by energy or entropy barriers, so that each configuration belongs to one and only one basin. A more detailed description of the energy landscape can be found in the cited literature.

At very low temperatures, the lifetime $θ_b$ of a finite-size system in a basin may be large compared to the particle vibration period τ. Then, the ergodicity is broken at times $t<θ_b$. A general discussion of systems with broken ergodicity may be found in the literature[20,21,22]. Above the glass transition temperature, this situation of broken ergodicity may be realized only in small systems. Due to the global nature of the energy landscape picture, the assumption of a long basin lifetime $θ_b$ is always violated in a sufficiently large system . An inter-basin step assumes a particle rearrangement in the system. The number 3N of dimensions of the configuration space, as well as the number of independently rearranging clusters, are proportional to the particle number N. Then, the frequency of inter-basin hopping is a global characteristic of the entire system, proportional to N.

A complementary local picture of liquids and glasses describes a small cluster randomly chosen in the system. Basic ideas of this picture were long ago discussed in the free volume theory [23],[24],[25],[26] (for a recent discussion of locality and more recent references see, for example, ref. [27],[28]). A particle of a cluster vibrates in a cage made by surrounding particles with a vibration period $\tau$; the amplitude of these vibrations is small compared to the distance between particles. The vibration motion conserves particle arrangement in the cluster. At times t~$\theta$, another component of thermal motion results in changes in cluster structure; this component is termed rearrangement motion. The time $\theta$ is then a local characteristic, the average lifetime of a stable arrangement in a small cluster. We consider the liquid in the temperature and pressure ranges where $\theta \gg \tau$.

The glassformer behavior suggests that rearrangement changes in the local structure are thermally activated[29],[30]. A fundamental suggestion is that global structural changes may be described in terms of local elementary events, thermally activated local rearrangements. Here, local means that while particle arrangement in a small rearranging cluster is significantly changed, particle displacements in the rest of the system, outside of the cluster, are small.

The link between the global and local pictures is established when a local rearrangement is identified as an elementary step in the inter-basin thermal motion. Rearrangements represent a form of thermal motion in the liquid; the fluctuating characteristic in this thermal motion is the local structure. Fluctuations of the local structure have their manifestations in both kinetic and thermodynamic properties of the system. Below, we consider relations between structural fluctuations and thermodynamic properties at different time scales. For equilibrium systems, relations between thermodynamic fluctuations and susceptibilities are well known (see, for example the textbook [31]). Those relations were used to estimate the contribution of structural fluctuations to susceptibilities of glassformer in conditions when this contribution is dominant (see, for example, [32],[33]). Below, we discuss structural and vibration contributions to susceptibilities, and give a straightforward derivation of these relations based on definition of basin thermodynamics. We then discuss an interpretation of these relations in terms of short and long time measurements. In the last section, we apply the derived relations to estimate the sensitivity of the local structure to changes of external parameters.

**2. Statistical mechanics of a glassformer in terms of basins**

In equilibrium[34], the probabilities to find the system in a small element of the configuration space are given by the Gibbs ensemble. By summing up the probabilities of all configurations belonging to a basin **a**, one finds the probability W(**a**) to find the system in this basin. Here, **a** is the basin label. One may interpret [33] the statistical ensemble as representing a large body made of identical N-particle systems. In this interpretation, different parts of the body occupy different basins, and W(**a**) is the probability to find a part in the basin **a**. For an ergodic system, W(**a**) also describes the fraction of time a part spends in this basin. General theorems of statistical mechanics justify the use of the ensemble $W_{eq}(\mathbf{a})$ to calculate equilibrium properties, including fluctuation characteristics.

In the Gibbs ensemble, the probability dW(q) to find a system of classical particles in an element Dq of the configuration space is

$$dW(q) = e^{\frac{F-H(q)}{k_B T}} Dq,$$
$$F = -k_B T \ln Z(T,V),$$
$$Z = \int e^{-\frac{H(q)}{k_B T}} Dq$$

(1)

Here, q are the particles coordinates, H(q) the microscopic Hamiltonian of the system, and F(T,V) the Helmholtz free energy. It is assumed that physically equivalent configurations, differing only by permutations of identical particles, are counted as one configuration; technically, this may be accounted for by dividing the result of independent integration over all particle coordinates by the number of corresponding permutations. Equilibrium (or long time-scale) pressure P, internal energy U, compressibility $\beta$, heat capacity C, and other thermodynamic characteristics are related to the thermodynamic potential F(T,V) by standard thermodynamic formulas.

As assumed in the literature (see cited references, and discussion in the next section), results of a short-time measurement are represented by the thermodynamics of a system confined in a basin. This basin thermodynamic is defined [16] by the basin free energy $F_a$

$$F_a(T,V) = -k_B T \ln Z_a \quad . \qquad 2$$

$$Z_a = \int_{q \in a} e^{-\frac{H(q)}{k_B T}} Dq , \qquad 3$$

where $a$ labels the basin. The basin partition function $Z_a$ is a sum over all intra-basin configurations $q \in a$, and $F_a$ is the basin free energy. The probability $W(a)$ to find the system in this basin is

$$W(a) = e^{\frac{F - F_a}{k_B T}},$$

$$Z = \sum_a e^{-\frac{F_a}{k_B T}} = e^{-\frac{F}{k_B T}} \quad . \qquad 4$$

In (4), the partition function Z(T,V) is rewritten in terms of basin contributions; the sum goes over the set of all physically different basins. The basin free energy $F_a$ plays the role of the effective Hamiltonian[35] for the basin label $a$ as a fluctuating variable.

One finds the basin thermodynamic characteristics by using the thermodynamic formulas with the basin free energy $F_a$ as a thermodynamic potential. For example, the basin thermal expansion coefficient $\alpha_p$ for a system that occupies, at the time of measurement, the basin a (as explained in the next Section, this actually means a narrow basin set including a), is

$$V\alpha_{p,a} = \left(\frac{\partial V}{\partial T}\right)_{P,a} = -\frac{\left(\frac{\partial P(a)}{\partial T}\right)_V}{\left(\frac{\partial P(a)}{\partial V}\right)_T} = -\frac{\frac{\partial^2 F_a}{\partial V \partial T}}{\frac{\partial^2 F_a}{\partial V^2}}.$$

A basin thermodynamic characteristic X(a) as a function of time fluctuates due to inter-basin (rearrangement) motion. At large times the system becomes ergodic. The large time average short-time value <X(a)> can be calculated as ensemble average

$$<X(a)> = \sum_a X(a)W(a).$$

Both the average short time values and the equilibrium values of observables are characteristics of the thermodynamic equilibrium.

**3. Inter-basin fluctuations**

At time scales $t \gg \theta$ of a long-time experiment the liquid approaches thermodynamic equilibrium; the structural equilibration mechanism is inter-basin (rearrangement) motion. At equilibrium, the rearrangement motion may be described as an unbiased random flight of the system in the basin set {a}, with basin label a(t) as a random variable. The relaxation kinetics, and the correlation of fluctuations in time are determined by the distributions of both energy minima and barrier energies, while the equilibrium probability to find the system in a basin is universally defined by the Gibbs distribution. Configurations with energies close to the top of the barrier have a negligible statistical weight at equilibrium, and correspondingly a negligible contribution to basin probability and to thermodynamics; the energy barriers only determine the time-scale at which a representative basin ensemble may be sampled. The universal Gibbs distribution for basin probabilities leads to universal relations between fluctuation characteristics and susceptibilities derived below.

Inter-basin fluctuations change the local structure; the thermodynamic manifestations of these changes are fluctuations of P(t), U(t), and other short-time characteristics. In the statistical mechanics of equilibrium systems, there is a general scheme[13] relating the fluctuation characteristics (binary correlation functions) to the susceptibilities (response functions). Below, we derive similar formulas that relate the magnitudes of inter-basin fluctuations to the differences between the average short-time and the equilibrium values of thermodynamic observables.

Consider a basin thermodynamic quantity X(a). The equilibrium value of X(a) describes the average of a long time (equilibrium) measurement of this quantity for a given part of the system. We use the notation X for this equilibrium value of X(a), and reserve the brackets <X> for the basin ensemble average

$$<X> = \sum_a X(a) W(a) = \sum_a X(a) e^{\frac{F-F_a}{k_b T}} .\qquad 7$$

For pressure P and internal energy U, the equilibrium values coincide with the ensemble-averages of short-time values. For the pressure P, and basin pressure P(a), defined as P=-($\partial$F/$\partial$V)$_T$, P(a)=-($\partial F_a/\partial V$)$_T$, one finds

$$P = -\frac{\partial}{\partial V}(-k_B T \ln \sum_a e^{-\frac{F_a}{k_B T}}) = \sum_a P(a) W(a) = \langle P(a) \rangle .\qquad 8$$

For the heat capacity $c_v$, thermal expansion coefficient $\alpha_p$, isothermal compressibility $\beta_T$ and the bulk modulus $K=1/\beta_T$, and other thermodynamic susceptibilities, the average short-time values differ from corresponding equilibrium values. The definitions, and measurements, of these quantities assume small changes in temperature and/or pressure. At a long time-scale, a change in temperature or pressure results in structural relaxation in the system, while the short time thermodynamics excludes contributions to susceptibilities coming from these structural changes. For the bulk modulus $K= - V(\partial P/\partial V)_T$, $K(a)= - V(\partial P(a)/\partial V)_T$, one obtains from (8)

$$K = -V \sum_a \left(\frac{\partial P(a)}{\partial V}\right)_T W(a) - \frac{V}{k_B T} \sum_a [P(a) - P]^2 W(a) . \qquad 9$$

K is the long-time, equilibrium characteristic, while the first term on the right is the average short-time bulk modulus. Formula (9) may be rewritten as

$$\left\langle [\delta P(a)]^2 \right\rangle = -\frac{k_B T}{V}[K - \langle K(a)\rangle] = -\frac{k_B T}{V} \Delta K . \qquad 10$$

Here, we introduced the notation $\delta X(a)$ for the deviation of a basin characteristics from its basin ensemble average, and $\Delta X$ for the difference between the equilibrium value X and the average short-time value $<X(a)>$ of the same characteristic:

$$\delta X(a) = X(a) - <X(a)>;$$
$$\Delta X = X - \sum_a X(a) W(a) = X - \langle X(a) \rangle . \qquad 11$$

Similarly to the derivation of (9), one obtains for the thermal coefficient of pressure $(\partial P/\partial T)_V$

$$\langle \delta P(a) \delta U(a) \rangle = k_B T^2 \Delta\left(\frac{\partial P}{\partial T}\right)_V ; \qquad 12$$

the internal energy U is defined by

$$U = F + TS = F - T\left(\frac{\partial F}{\partial T}\right)_V = \langle U(a) \rangle;$$

$$U(a) = F_a + TS(a) = F_a - T\left(\frac{\partial F_a}{\partial T}\right)_V \quad \mathbf{13}$$

By using the thermal expansion coefficient $\alpha_P = (1/V)(\partial V/\partial T)_P$ related to $(\partial P/\partial T)_V$ by $K\alpha_P = (\partial P/\partial T)_V$, $K(a)\alpha_P(a) = (\partial P(a)/\partial T)_V$, one rewrites (12) as

$$\langle \delta P(a) \delta U(a) \rangle = k_B T^2 \Delta(K\alpha_P) . \quad \mathbf{14}$$

The structural contribution to the thermal expansion coefficient was recently calculated by Stillinger and Debenedetti (ref.[16]). For the heat capacity $C_V$, one obtains

$$C_V = T\left(\frac{\partial S}{\partial T}\right)_V = -T\frac{\partial}{\partial T}\left(\frac{\partial F}{\partial T}\right) = \langle C_V \rangle + \frac{1}{k_B T^2}\langle [\delta U(a)]^2 \rangle. \quad \mathbf{15}$$

This relation may be rewritten as

$$\langle [\delta U(a)]^2 \rangle = k_B T^2 V \Delta c_V , \quad \mathbf{16}$$

where $c_v = C_V/V$.

Formulas (10),(12),(14),(16) give the fluctuations-to-susceptibilities relations for inter-basin fluctuations. Thermal equilibrium at the long time scale is a necessary condition for validity of these relations. This condition may appear challenging for fragile glassformers in the vicinity of their glass transition because of a wide range of relaxation times in these systems[8]. The fraction of the fragile glassformer that has the relaxation times larger than the time of long-time experiments increases upon supercooling. When this fraction become significant, one arrives at a crossover from short to long time behavior; when the fraction of structurally frozen clusters becomes dominant, only short-time regime may be realized.

The formulas derived in this Section relate independently measurable quantities, and thus provide a way to directly test the suggested mechanism of inter-basin fluctuations. Relatively small systems may be studied by the computer simulation method. New opportunities to experimentally study pressure fluctuations are provided by optical measurements involving chromophores (spectral hole dynamics, single molecule spectroscopy) in small systems imbedded in a rigid matrix. Indirectly, one may test predictions that are based on the derived relations. A prediction of relative sensitivity of the structure to changes in thermodynamic parameters is discussed in the next Section.

## *4. Time scale for short-time kinetics.*

For a system with a long lifetime $\theta_b \gg \tau$ in a basin, ergodicity at times $t < \theta_b$ is broken. A general discussion of short time properties for such a system may be found in the cited ref. 20-22. Average value of the internal energy U, pressure P, volume V, and other quantities over a time t, $\tau \ll t \ll \theta_b$, may be approximated as equilibrium properties for a system when it is confined in the basin. In a large system, however, the inter-basin fluctuations of pressure P(t) and other thermodynamic characteristics have a high frequency proportional to the macroscopic particle number N. We present here arguments that the basin thermodynamics may be applied to short-time measurements at times $t < \theta$; the time $\theta$ is a local characteristic independent from the particle number N in the system.

At a short time-scale $t \ll \theta$, rearrangements take place in a small fraction $\sim t/\theta$ of the material, while in the rest of the material the structure is conserved. For additive thermodynamic quantities (volume, pressure, internal energy), contributions of different parts of the system are proportional to parts volume. Then, the relative contribution of those part where rearrangements took place is proportional to the small volume fraction ($\sim t/\theta \ll 1$) of this parts, and may be neglected. With this accuracy, one may describe the results of short-time measurement by the theoretical model of a system confined to a basin. The system actually changes basins during the short-time measurement, but the local structure for the basins visited by the system during the time of measurement coincide in most of the volume. The contribution of this structurally conserved part of the system to short-time thermodynamics coincides, up to small terms of the order of $t/\theta$, with the contribution of the system confined to any one of the visited basins. With this accuracy, the derived relations are between <u>experimentally observable</u> fluctuation and thermodynamic characteristics.

## 5. Applicability to supercooled liquids

It is well known that good glassformer may remain in a metastable supercooled state for a very long time, behaving at this stability time similar to stable systems. One may then approximate the basin ensemble for these supercooled liquids by the Gibbs ensemble. This approximation neglects the slow evolution of the metastable ensemble towards stable equilibrium state; this evolution is considered in the nucleation theory[36]. A metastable liquid remains in an approximately steady state only until the thermal inter-basin motion brings the system to a basin corresponding to crystalline clusters with the size about or larger than the critical nucleus size. Those states and basins will be referred to as crystalline. To model the metastable state, one considers a truncated configuration space that includes statistically significant structural states of the glassformer at the time when the metastable state is still relatively stable, and excludes crystalline states. Note that the probability to find an equilibrium liquid in a crystalline state is negligibly small, so that the statistical mechanics for this system in the truncated configuration space coincides with that in the full configuration space.

The formulas derived in above Sections are based on a Gibbs statistics. In the approximation of Gibbs statistics in the truncated basin set, these formulas may be applied to supercooled liquids. This application, however, assumes that equilibration in the truncated configuration space takes place at time scale when the system is steady. Changes in the metastable ensemble due to the growth of subcritical crystalline nuclei in the liquid matrix are small only at time-scales short compared to the lifetime of the metastable state. By using the approximation of Gibbs equilibrium in the truncated configuration space, we limit the applicability of our theory to supercooled states characterized by lifetimes much larger than the structural relaxation and equilibration times, so that the metastable system can sample a representative ensemble of local structures on the truncated set of basins. This sets a limit to the concept of a steady metastable state with properties independent from the thermal history of the system.

**6. Controlling parameters in supercooling**

The aim of this Section is to compare changes in the local structure caused by temperature and volume (or pressure) changes. Recent observations[37,4] have shown little difference between the increases of viscosity $\eta$ and local relaxation time $\theta$ upon cooling the liquid at constant pressure or constant volume. The conclusion of these studies is that the line $\theta,\eta$=const in the thermodynamic P,T plane is much steeper than the isochore V=const.

Generally, any function Y(T,V) of the thermodynamic state defines a direction in the (T,V) thermodynamic plane of constant Y(T,V): the condition Y(T,V)=const in differential form is

$$dV = p_Y dT, \quad p_Y = -\frac{\left(\frac{\partial Y}{\partial T}\right)_V}{\left(\frac{\partial Y}{\partial V}\right)_T}. \qquad 17$$

The parameter $p_Y$ determines the relative sensitivity of Y to volume and temperature changes: a temperature change dT results in the same change dY as the volume change $dV = p_Y dT$. The pressure change $dP_Y = K dV/V$ that results in the isothermal volume change dV, equals

$$dP_Y = -\frac{K}{V} dV, \qquad 18$$

$K = -V(\partial P/\partial V)_T$ is the equilibrium bulk modulus. For example, the choice Y=V gives the isochoric pressure change $dP_V$

$$dP_Y = K \alpha_P dT, \qquad 19$$

$\alpha_P$ is the thermal expansion coefficient. This relation gives a thermodynamic "compatibilizer" to compare changes in pressure and temperature.

Changes in thermodynamic and kinetic properties reflect both changes at constant structure (changes in the vibration amplitude, density changes at constant structure (the mechanism better known in crystals), and structural changes. The general cause of structural changes is that, according to the general principles of statistical mechanics, the statistical weight of basins with lower free energies $F_a$ increases upon cooling; upon isothermal compression, the probability increases to find the system in a basin with more dense packing. The structure of the system is statistically described by the basin probability distribution W(a). When the temperature and pressure change, the probability W(a) to find the system in a basin a, and thus the structure, changes. For the relative change of the probability W(a) one finds

$$\frac{\delta W(a)}{W(a)} = \frac{1}{W(a)}\left[\left(\frac{\partial W(a)}{\partial T}\right)_V dT + \left(\frac{\partial W(a)}{\partial V}\right)_T dV\right] = -\frac{1}{k_B T}\left[\frac{U - U(a)}{T} dT + [P - P(a)] dV\right]. \qquad 20$$

For one basin, the condition $\delta W(a)=0$ determines the volume change $dV_a$ that compensates for the action of a temperature change dT:

$$dV_a = -\frac{U-U(a)}{T(P-P(a))}dT. \qquad 21$$

The compensating volume change $dV_a$ depends on the basin chosen. One cannot expect to simultaneously compensate the probability changes $\delta W(a)$ for all basins. In other words, there are structure changes for any change of temperature and volume or pressure. For a given characteristic that depends on structure, for example for the viscosity $\eta$, one can compensate the changes due to temperature increase dT by changes due to appropriate increase of pressure dP. The ratio dP/dT may be estimated by comparing structure changes caused by dT and dP. The ensemble average of the relative change $\delta W/W$ defined in formula (20) vanishes:$<\delta W/W>=0$. One considers then the ensemble average of $(\delta W/W)^2$. A non-negative quadratic form Q, proportional to this ensemble average is

$$Q(T,V) = \sum_a \left[\frac{\delta U(a)}{T}dT + \delta P(a)dV\right]^2 W(a). \qquad 22$$

Instead of volume change dV, one uses the corresponding isothermal pressure change $dP=-KdV/V$, to bring the quadratic form Q to the form

$$Q(T,V) = A_{TT}(dT)^2 - 2A_{TP}dTdP + A_{PP}(dP)^2, \qquad \mathbf{23}$$

with

$$A_{TT} = \frac{1}{T^2}\langle \delta U(a)^2 \rangle;$$

$$A_{TP} = \frac{V}{TK}\langle \delta U(a)\delta P(a)\rangle; \quad\quad 24$$

$$A_{PP} = \frac{V^2}{K^2}\langle (\delta P(a))^2 \rangle$$

With the help of (10),(12), and (16), one writes the coefficients in terms of thermodynamic characteristics:

$$A_{TT} = k_B V \Delta c_V,$$

$$A_{TP} = \frac{1}{K} k_B T V \Delta(K\alpha_P), \quad\quad 25$$

$$A_{PP} = k_B T V \frac{|\Delta K|}{K^2}.$$

The condition Q=const determines an ellipse in the dP-dT plane. As a measure of compensation of structure changes due to dT and dP is given by the ratio

$$s = \frac{(A_{PT})^2}{A_{TT}A_{PP}} \quad\quad 26$$

A complete compensation corresponds to s=1. A rather rough estimation of the coefficients in Q using extrapolation of the data on atactic polystyrene[38],[39],[40] at P~1 bar, T=$T_g$=150C gives s~1/500. This small value shows that the correlation of fluctuations of the energy U and pressure P is small, and one can not compensate structure changes caused by temperature change with those caused by a change of the pressure.

A measure of the relative sensitivity of the structure to temperature and pressure is given by the ratio

$$q = \sqrt{\frac{A_{TT}}{A_{PP}}} = K^2 \frac{\Delta(c_V)}{T|\Delta K|}; \quad\quad 27$$

from the mentioned data on polystyrene one gets

$$q \sim 15 \frac{bar}{°K} \qquad 28$$

The coefficient q is larger but close to the isochoric coefficient $(\partial P/\partial T)_V = K\alpha_P \approx 10$ bar/K. The accuracy of this estimation is low because the experimental data are for the glass transition range in the thermodynamic plane; as already mentioned, application of the theory to fragile glassformers close to the glass transition temperature imposes challenging conditions for the equilibration time. More accurate experimental data are needed to make quantitative predictions.

The slope of the glass transition line (line of constant viscosity $\eta = 10^{13}$ p) of polystyrene gives a substantially larger coefficient $p_\eta = (\partial \eta/\partial T)/(\partial \eta/\partial P) \approx 40$ bar/Kelvin. Some part of this coefficient may be explained by the fact that viscosity and relaxation time increase upon cooling even in a system with energy barriers not changing. At low temperatures, the data on viscosity of the liquid are traditionally fitted by the formula $\eta = \eta_0 \exp(\Delta/k_B T)$, with $\eta_0 \sim 0.01$ poise and the effective excitation energy $\Delta(T,P)$ a function of temperature and pressure. Similar formulas may be written for relaxation time and other kinetic characteristics. At the glass transition, by convention, $\eta = 10^{13}$ poise, hence $\Delta/k_B T_g \approx 35$. The constant viscosity line in the (P,T) thermodynamic plane yields the relation

$$k_B T d(\ln \eta) = \left( -\frac{\Delta}{T} + \frac{\partial \Delta}{\partial T} \right) dT + \frac{\partial \Delta}{\partial P} dP = 0, \qquad 29$$

or, for the coefficient $p_\eta$

$$p_\eta = -\frac{\frac{\partial \eta}{\partial T}}{\frac{\partial \eta}{\partial P}} = \frac{\frac{\Delta}{T} - \frac{\partial \Delta}{\partial T}}{\frac{\partial \Delta}{\partial P}} \qquad 30$$

Strong glasses [6,7] are characterized by small changes of Δ(T,V) upon cooling, while in fragile glassformers Δ(T,V) increases significantly. One possible cause of this increase is the increase of barriers without structural changes, due to the increase in density. Another source of fragility is due to the dependence of the excitation energies on local packing; the structural changes upon cooling towards more ordered packing (usually also increasing the density) result in a sharp increase in Δ. Some details of both mechanisms are studied recently by Sastry[41]. We suggest (see discussion in [42]) that structural changes are the main cause of fragility in liquids studied in the cited papers [18,4]. Then, as a first approximation, one may neglect excitation energy changes at constant structure. As shown above, any change in thermodynamic parameters T and P changes the structure, and compensation is impossible. One can, however, always compensate changes in viscosity, or in the apparent excitation energy Δ; this last condition define the coefficient $p_\Delta$

$$\frac{\partial \Delta}{\partial T} = -p_\Delta \frac{\partial \Delta}{\partial P} \qquad 31$$

One speculates that the sensitivity of the excitation energy to temperature and pressure is close to that of the structure, and $p_\Delta \approx q$, where q is defined in (27). From this formula and the definition $m(T) = \partial \log\eta / \partial(T_g/T)$, $m(T_g)$ being the Angell fragility characteristic [6,7], one can rewrite (30) as

$$p_\eta = p_\Delta \frac{mMk_B T_g}{mMk_B T_g - \Delta} \qquad 32$$

Here, $M = \ln 10 \approx 2.3$. At temperatures close to the glass transition range, for fragile glassformers $mM < \Delta/k_B T_g$, and thus $p_\eta \approx p_\Delta$. When considered as a function of temperature and pressure, m decreases upon heating or lowering the pressure; for fragile glassformers, the denominator in formula (32) may become small, and correspondingly the coefficient $p_\eta > p_\Delta$ as observed.

**Acknowledgements**

We acknowledge financial support of this study by the Chemistry division of the NSF, by the Northwestern MRSEC, and NASA (Grant NAG-1932).